\documentclass[a4paper,12pt]{article}
\usepackage{amssymb}
\date{}
\title{One Loop Vacuum Energy in Non-parallel $\textrm{D}_1$-Branes:\\
A Path Integral Formulation}
\author{A. Jahan\\Department of Physics, Amirkabir University of Technology, Tehran, Iran\\jahan@aut.ac.ir}
\begin{document}
\maketitle
\begin{abstract}
We derive the one loop vacuum energy of the bosonic string theory in a system of non-parallel $\textrm{D}_1$-branes using the path integral method.
\end{abstract}
\section*{\large 1\quad Introduction}
The importance of D-branes (Dirichlet branes)  in string theory relies on the fact that such an extended objects may play a vital role in non-perturbative description of the relationships between the different theories of strings [1, 2, 3]. In particular they play a very specific role in describing the duality symmetries [4, 5]. On the other hand it is pointed out that the tension between the branes reveals itself through the vacuum amplitude in a
system of parallel D-branes [6, 7].\\
In this letter we consider a non-parallel system of $\textrm{D}_1$\textmd{-}branes and derive its one loop amplitude for the bosonic strings upon applying the path integral method as considered in [6, 7, 8]. Such a system is studied earlier by utilizing the harmonic oscillator formalism for the strings attached to the system through their end points [9, 10, 11]. We start our analyze by specifying the position of the branes constructing the set up, in section 2. Then, after a quick review of the path integral formulation of the one loop vacuum energy in section 3, we look for the partition functions of the degrees of freedom satisfying the Neumann-Neumann (NN), Dirichlet-Dirichlet (DD) and mixed boundary conditions in following subsections. Finally in section 4, we give the final result for the one loop vacuum amplitude. \\
As a final remark, let us notify that in this work we have assumed the Einstein summation rule for repeated indices to hold just for the Greek indices. We have also assumed the Euclidean signature for both of the world-sheet and space-time manifolds.
\section*{\large 2\quad Non-parallel $\textrm{D}_1\textmd{-}\textrm{D}_1$ Brane Set Up}
Let us suppose the position of first $\textrm{D}_1$-brane to be
\begin{equation}\label{1}
X^i(0,\tau)=0,\quad\quad i=2,...,25
\end{equation}
and the second one to be located at
{\setlength\arraycolsep{2pt}
\begin{eqnarray}\label{2-3}
X^2(\pi,\tau)\cos\alpha&=&X^1(\pi,\tau)\sin\alpha\\
X^r(\pi,\tau)&=&l_r
\end{eqnarray}}
where $r=3,...,25$. Then, the conditions satisfied by the ends of an open string at the boundaries, imposed by the classical equations of motion, read [9, 10, 11]
{\setlength\arraycolsep{2pt}
\begin{eqnarray}\label{4-6}
\partial_\sigma X^1(0,\tau)&=&0\\
X^2(0,\tau)&=&0\\
\partial_\sigma X^1(\pi,\tau)\cos\alpha&=&-\partial_\sigma X^2(\pi,\tau)\sin\alpha
\end{eqnarray}}
\section*{\large 3\quad One Loop Vacuum Energy: Path Integral Formulation}
For the bosonic sigma model action
\begin{equation}\label{7}
S=\frac{g}{2}\int d^2\sigma\,\partial_\sigma X^\mu\partial_\sigma X^\mu=-\frac{g}{2}\int d^2\sigma\,X^\mu\Box X^\mu+boundary\,\,term
\end{equation}
the partition function is defined as
\begin{equation}\label{8}
\mathcal Z=\int DX^\mu e^{-S[X]}
\end{equation}
By expanding the typical degree of freedom $X^\mu$ in terms of the fluctuating filed around the on-shell
degrees of freedom $X^\mu_c$, which satisfies the classical equations of motion, as $X^\mu=X^\mu_c+\xi^\mu$, one is left with
\begin{equation}\label{9}
\mathcal Z=\int D\xi^\mu e^{-S[\xi]}
\end{equation}
where we have set $S[X_c]=0$, by demanding an appropriate boundary condition to be satisfied by $X_c$. Off-shell, the fluctuating field $\xi^\mu$ is assumed to nullify the boundary term of equations (4), (5) and (6), by choosing appropriate eigen-modes in terms of which the field $\xi^\mu$ is expanded. The one loop vaccum amplitude (or free energy) is defined to be [6-11]
{\setlength\arraycolsep{2pt}
\begin{eqnarray}\label{10}
\mathcal A&=&\int_0^\infty \frac{ds}{s}Tre^{-sH}=\int_0^\infty \frac{ds}{s}\mathcal Z(s)\\
&=&\int_0^\infty \frac{ds}{s}\int D\xi^\mu e^{-S[\xi]}\nonumber
\end{eqnarray}}
provided that $\xi^\mu(\tau+s,\sigma)=\xi^\mu(\tau,\sigma)$. Now, the above expression is our starting point to digress from the Hamiltonian formalism (as appeared in [9, 10, 11]) and to purse the path integral formulation of the problem.
\section*{\small 3.1\quad Neumann-Neumann Boundary Condition}
For the degree of freedom $X^0$ satisfying the NN boundary condition we have
{\setlength\arraycolsep{2pt}
\begin{eqnarray}\label{11}
S_0=-\frac{g}{2}\int d^2\sigma\xi^0 \Box \,\xi^0=\frac{g}{2}\sum_{n^\prime\in\mathbb{N},m^\prime\in\mathbb{Z}}\sum_{n\in\mathbb{N},m\in\mathbb{Z}}\chi^0_{n^\prime m^\prime}\chi^0_{nm}\lambda_{nm}(u_{n^\prime m^\prime},u_{nm})
\end{eqnarray}}
and $\Box u_{nm}=-\lambda_{nm} u_{nm}$. The fluctuation in terms of the eigen-modes $u_{nm}=\cos n\sigma e^{i\omega_m\tau}$ (with $\omega_m=\frac{2\pi m}{s}$), which nullify the boundary term of the action, becomes
\begin{equation}\label{12}
\xi^0=\sum_{n\in\mathbb{N},m\in\mathbb{Z}}\chi^0_{nm}u_{nm}
\end{equation}
Then, by taking into account the following set of orthogonality relations
{\setlength\arraycolsep{2pt}
\begin{eqnarray}\label{13-16}
(u_{n^\prime m^\prime},u_{nm})&=&\frac{s}{2}\pi\delta_{n n^\prime}\delta_{m+m^\prime}\\
(u_{n^\prime m^\prime},u_{n0})&=&(u_{n^\prime 0},u_{nm})=0\\
(u_{0 m^\prime},u_{0m})&=&s\pi\delta_{m+m^\prime}\\
(u_{0 m^\prime},u_{nm})&=&(u_{n^\prime m^\prime},u_{0m})=0
\end{eqnarray}}
where we have defined
\begin{equation}\label{17}
(f,g)=\int_0^\pi d\sigma \int_0^s d\tau\, fg
\end{equation}
and bearing in mind that $\textbf{x}_0$ is real, we find for the action
{\setlength\arraycolsep{2pt}
\begin{eqnarray}\label{18}
S_0=\frac{1}{2}\pi sg\bigg(\sum_{m=1}^{\infty}\textbf{x}^{0\dagger}_{+,m}\textbf{M}_m\textbf{x}^0_{+,m}
+\frac{1}{2}\textbf{x}^{0 \,\scriptsize\textrm{t}}_{+,0}\textbf{M}_0\textbf{x}^0_{+,0}+2\sum_{m=1}^\infty|\chi^0_{0m}|^2\lambda^2_{0m}\bigg)
\end{eqnarray}}
with $[\textbf{M}_m]_{n^\prime n}=\delta_{n^\prime n}\lambda_{nm}$ and $\textbf{x}^{0\dagger}_{\pm,m}=(\bar{\chi}^0_{\pm1,m},\bar{\chi}^0_{\pm2, m},...)$.
In particular one obtains the partition function ($q=e^{-s}$) [6, 7]
{\setlength\arraycolsep{2pt}
\begin{eqnarray}\label{19}
Z_0(s)&=&\int D\xi^0e^{-S_0}\\\nonumber
&=&\prod_{m=1}^{\infty}\int_{-\infty}^{\infty}d(\textbf{x}^0_0,\textbf{x}^0_{+,m},\textbf{x}^{0 \dagger}_{+,m};\chi^0_{0m},\bar{\chi}^0_{0m},\chi^0_{00}) e^{-S_0}\\
&=&T\prod_{n=1}^{\infty}\frac{2}{\sqrt{sg}}\frac{1}{\sqrt{\lambda_{n0}}}
\prod_{m=1}^\infty\frac{2}{sg}\frac{1}{\lambda_{0m}}\prod_{n,m=1}^{\infty}\frac{4}{sg}\frac{1}{\lambda_{nm}}\nonumber\\
&=&T\sqrt{\frac{g}{2s}}q^{-\frac{1}{24}}\prod_{n=1}^\infty\frac{1}{1-q^{n}}\nonumber
\end{eqnarray}}
Here we have introduced the compact notation $d(x,y,...)=dxdy...$. Also, the following set of formulas are used
{\setlength\arraycolsep{2pt}
\begin{eqnarray}\label{20-22}
\frac{\sinh\pi{x}}{\pi{x}}&=&\prod_{n=1}^\infty\bigg(1+\frac{x^2}{n^2}\bigg)\\
\prod^{\infty}_{m=1}\frac{1}{a m^2}&=&\frac{\sqrt{a}}{2\pi}\\
\prod_{n=1}^\infty c&=&\frac{1}{\sqrt{c}}
\end{eqnarray}}
Note that integration over the zero mode $\chi^0_{00}$ yields the volume along the $X^0$ axis or the interval of time $T$, during which the interaction between two branes takes place.
\section*{\small 3.2\quad Mixed Boundary Condition}
For the fluctuations which satisfy the mixed boundary condition of Eqs.(4), (5) and (6), namely
{\setlength\arraycolsep{2pt}
\begin{eqnarray}\label{23-25}
\partial_\sigma \xi^1(0,\tau)&=&0\\
\xi^2(0,\tau)&=&0\\
\partial_\sigma \xi^1(\pi,\tau)\cos\alpha&=&-\partial_\sigma \xi^2(\pi,\tau)\sin\alpha
\end{eqnarray}}
we choose the eigen-modes to be $u^\alpha_{nm}=\cos\sigma n_\alpha e^{i\omega_m\tau}$ and $v^\alpha_{nm}=\sin\sigma n_\alpha e^{i\omega_m\tau}$, and expand them as
\begin{equation}\label{26}
\left\{ \begin{array}{ll}
\xi^1\\
\xi^2
\end{array} \right\}=\sum_{n,m\in\mathbb Z}\chi_{nm}
\left\{ \begin{array}{ll}
u^\alpha_{nm}\\
v^\alpha_{nm}
\end{array} \right\}
\end{equation}
with common eigen-value $\lambda^a_{nm}=n^2_a+\omega^2_m=(n+a)^2+\omega^2_m$. The number $a=\frac{\alpha}{\pi}$ takes the values $0 \leq a\leq1$ and the eigen-modes fulfill the following set of relations
{\setlength\arraycolsep{2pt}
\begin{eqnarray}\label{27-30}
(u^\alpha_{n^\prime m^\prime},u^\alpha_{nm})&=&\frac{s}{2}\pi \Bigg(\delta_{n n^\prime}+\frac{(-1)^{n+n^\prime}}{\pi}
\frac{\sin2\pi a}{n+n^\prime+2a}\Bigg)\delta_{m+m^\prime}\\
(u^\alpha_{n^\prime m^\prime},u^\alpha_{n0})&=&(u^\alpha_{n^\prime 0},u_{nm})=0\\
(v^\alpha_{n^\prime m^\prime},v^\alpha_{nm})&=&\frac{s}{2}\pi \Bigg(\delta_{n n^\prime}-\frac{(-1)^{n+n^\prime}}{\pi}
\frac{\sin2\pi a}{n+n^\prime+2a}\Bigg)\delta_{m+m^\prime}\\
(v^\alpha_{n^\prime m^\prime},v^\alpha_{n0})&=&(v^\alpha_{n^\prime 0},v^\alpha_{nm})=0
\end{eqnarray}}
 Now the choice
\begin{equation}\label{26}
\Phi=\frac{1}{\sqrt 2}
\left(\begin{array}{ccc}
\xi^1\\
\xi^2\\
\end{array}\right)
\end{equation}
diagonalizes the action. So, by following the same steps which led to the equation (18) and noting that $\lambda^a_{-nm}=\lambda^{-a}_{nm}$ we get
{\setlength\arraycolsep{2pt}
\begin{eqnarray}\label{31}
S_2&=-&\frac{g}{2}\int d^2\sigma\,\Phi^{{\scriptsize\textrm{t}}} \textbf{D}\Phi
=\frac{1}{2}\pi sg\sum_{m=1}^{\infty}\bigg(\textbf{x}^{\dagger }_{+,m}\textbf{M}^a_m\textbf{x}_{+,m}+\textbf{x}^{\dagger }_{-,m}\textbf{M}^{-a}_m\textbf{x}_{-,m}\bigg)\\\nonumber
&+&\frac{1}{4}\pi sg\bigg(\textbf{x}^{\scriptsize\textrm t}_{+,0}\textbf{M}^a_0\textbf{x}_{+,0}+\textbf{x}^{\scriptsize\textrm t}_{-,0}\textbf{M}^{-a}_0\textbf{x}_{-,0}
+2\sum_{m=1}^\infty|\chi_{0m}|^2\lambda^a_{0m}+\sum_{m=1}^\infty\chi_{00}^2\lambda^a_{00}\bigg)
\end{eqnarray}}
where we have introduced the $2\times2$ matrix $\textbf{D}=\textbf{1}\Box$. We have also assumed $[\textbf{M}^a_m]_{n n^\prime}=\lambda^a_{nm}\delta_{nn^\prime}$. Therefore, we find the partition function as
{\setlength\arraycolsep{2pt}
\begin{eqnarray}\label{32}
Z_2(s)&=&\prod_{m=1}^{\infty}\int^{\infty}_{-\infty} d(\textbf{x}_{-,0},\textbf{x}_{+,0},\textbf{x}_{+,m},\textbf{x}^\dagger_{+,m},\textbf{x}_{-,m},\textbf{x}^\dagger_{-,m};\chi_{0m},\bar{\chi}_{0m},{\chi}_{00}) e^{-S_2}\\\nonumber
&=&\Bigg(\prod_{n=1}^{\infty}\frac{2}{\sqrt{sg}}\frac{1}{\sqrt{\lambda^a_{n0}}}\prod_{n,m=1}^{\infty}\frac{4}{sg}\frac{1}{ \lambda^a_{nm}}\Bigg)
\Bigg(\prod_{n=1}^{\infty}\frac{2}{\sqrt{sg}}\frac{1}{\sqrt{\lambda^{-a}_{n0}}}\prod_{n,m=1}^{\infty}\frac{4}{sg}\frac{1}{\lambda^{-a}_{nm}}\Bigg)\\\nonumber
&\times&\Bigg(\frac{2}{\sqrt{sg}}\frac{1}{\sqrt{\lambda^a_{00}}}\prod_{m=1}^{\infty}\frac{4}{sg}
\frac{1}{\lambda^a_{0m}}\Bigg)\\\nonumber
&=&\frac{q^{\frac{a}{2}(1-a)-\frac{2}{24}}}{1-q^a}\prod_{n=1}^\infty\frac{1}{1-q^{n-a}}\frac{1}{1-q^{n+a}}\nonumber
\end{eqnarray}}
where we have utilized the well-known formula
{\setlength\arraycolsep{2pt}
\begin{eqnarray}\label{33}
\sum^{\infty}_{n=1}(n+a)=\frac{1}{24}-\frac{1}{2}\bigg(a+\frac{1}{2}\bigg)^2
\end{eqnarray}}
\section*{\small 3.3\quad Dirichlet-Diriclet  Boundary Condition}
To evaluate the contribution arising from the degrees of freedom satisfying the DD boundary condition, i.e. the coordinates $X^r$ with $3\leq r$, we expand the corresponding fluctuations as
\begin{equation}\label{34}
\xi^r=\frac{\sigma}{\pi}l_{\,r}+\sum_{n\in\mathbb{N}^+,m\in\mathbb{Z}}\chi^r_{nm}v_{nm}
\end{equation}
where $\mathbb{N}^+=\mathbb{N}-\{0\}$. With the eigen-modes $v_{nm}=\sin\sigma ne^{i\omega_m\tau}$, we find the action to be
{\setlength\arraycolsep{2pt}
\begin{eqnarray}\label{35}
S_{23}=\frac{1}{2}\pi sg\sum^{25}_{r=3}\bigg(\sum_{m=1}^{\infty}\textbf{x}^{r \dagger}_{+,m}\textbf{M}_m\textbf{x}^r_{+,m}
+\frac{1}{2}\textbf{x}^{r\,\scriptsize\textrm t}_{+,0}\textbf{M}_0\textbf{x}^r_{+,0}\bigg)+\frac{1}{2\pi}sg\sum^{25}_{r=3}l^2_r
\end{eqnarray}}
The last term of the above expression arises from the boundary term
\begin{equation}\label{36}
\frac{g}{2}\int^s_0 d\tau \xi^r(\pi,\tau)\partial_\sigma \xi^r(\pi,\tau)=\frac{1}{2\pi}sgl_{\,r}^{\,2}
\end{equation}
Now we have the partition function as
{\setlength\arraycolsep{2pt}
\begin{eqnarray}\label{37}
Z_{23}(s)&=&\prod_{r=3}^{25}\prod_{m=1}^\infty\int^{\infty}_{-\infty} d(\textbf{x}^r_{+,0},\textbf{x}^r_{+,m},\textbf{x}^{r\dagger}_{+,m})
e^{-S_{23}}\\\nonumber
&=&e^{-\frac{1}{2\pi}{s}g\sum_{r=3}^{25}l^2_r}\Bigg(\prod_{n=1}^\infty\frac{2}{\lambda_{n0}\sqrt{g{s}}}
\prod_{n,m=1}^\infty\frac{4}{sg}\frac{1}{\lambda^2_{nm}}\Bigg)^{23}\\\nonumber
&=&q^{\frac{1}{2\pi}gY^2}\Bigg(q^{-\frac{1}{24}}\prod^\infty_{n=1}\frac{1}{1-q^{n}}\Bigg)^{23}\nonumber
\end{eqnarray}}
with $Y^2=\sum_{r=3}^{25}l^2_r $.
\section*{\large 4\quad Final Result for Vacuum Amplitude}
Therefore, having at hand the contributions of several degrees of freedom satisfying different boundary conditions, we find the final result for the partition function
{\setlength\arraycolsep{2pt}
\begin{eqnarray}\label{38}
\mathcal{Z}(s)&=&Z_0(s)Z_2(s)Z_{23}(s)Z_{gh}(s)\\\nonumber
&=&T\sqrt{\frac{g}{2s}}\frac{q^{\frac{1}{2\pi}gY^2-1-\frac{a}{2}(a-1)}}{1-q^a}
\prod_{n=1}^\infty(1-q^{n})^{-22}(1-q^{n+a})^{-1}(1-q^{n-a})^{-1}
\end{eqnarray}}
which yields the one loop amplitude as
\begin{equation}\label{39}
\mathcal {A}=T\int_0^\infty\frac{ds}{s}\sqrt{\frac{g}{2s}}\frac{q^{\frac{1}{2\pi}gY^2-1-\frac{a}{2}(a-1)}}{1-q^a}
\prod_{n=1}^\infty(1-q^{n})^{-22}(1-q^{n+a})^{-1}(1-q^{n-a})^{-1}
\end{equation}
where the contribution of the ghost and anti-ghost fields is taken into account by setting [6]
\begin{equation}\label{40}
Z_{gh}=\frac{g}{2s}Z^{-2}_0
\end{equation}
Now, Eq.(40) stands as the main achievement of this present work. The result which was derived earlier in [9, 10, 11] on employing the Hamiltonian formalism.
\section*{\large Acknowledgments}
Author would like to thank Dr. D. Kamani for helpful comments and fruitful conversations.

\section*{\large References}
[1] \hspace{0.2cm}C. M. Hull, P. K. Townsend, Nucl. Phys. \textbf{B438} (1995),109.\\\
[2] \hspace{0.2cm}A. Sen, Nucl. Phys. \textbf{B450} (1995), 103.\\\
[3] \hspace{0.2cm}E. Witten, Nucl. Phys. \textbf{B443} (1995), 85.\\\
[4] \hspace{0.2cm}J. Polchinski, S. Chaudhuri and C.V. Johnson, arXiv: hep-th/9602052.\\\
[5] \hspace{0.2cm}J. Polchinski, Phys. Rev. Lett. \textbf{75} (1995), 4724.\\\
[6] \hspace{0.2cm}C. Acatrinei, Nucl. Phys. \textbf{B539} (1999) 513. \\\
[7] \hspace{0.2cm}C. Acatrinei, R. Iengo, Phys. Lett. \textbf{B482} (2000) 420.\\\
[8] \hspace{0.2cm}J. Ambjorn, Y. M. Makeenko, G. W. Semenoff and R. J. Szabo, JHEP 0302 (2003) 026. \\\
[9] \hspace{0.2cm}A. Matusis, Int. J. Mod. Phys. \textbf{A14} (1999) 1153.\\\
[10] \hspace{0.2cm}H. Arfaei, M. M. Sheikh-Jabbari, Phys. Lett. \textbf{B394} (1997) 288.\\\
[11] \hspace{0.2cm}T. Kiato, N. Ohta and J. Zhaou, Phys. Lett. \textbf{B428} (1998) 68.

\end{document}